\documentclass[a4paper]{article}

\usepackage{graphicx}
\usepackage{amsmath,amssymb}
\usepackage{authblk}
\newcommand{\affiliation}[1]{\affil{#1}}

\begin{document}

\title{An Intrinsic Way to Control E-Sail Spin}

\author{Pekka Janhunen\footnote{Research Manager.}
and Petri Toivanen\footnote{Senior Scientist.}}
\affiliation{Finnish Meteorological Institute, POB--503, Helsinki, FI-00101, Finland}

\maketitle

\begin{abstract}
We show that by having the auxtethers made partly or completely of
conducting material and by controlling their voltages, it is possible
to control the spin rate of the electric solar wind sail by using the
electric sail effect itself. The proposed intrinsic spin rate control
scheme has enough control authority to overcome the secular change of
the spin rate due to orbital Coriolis effect.
\end{abstract}

\section*{Nomenclature}

\noindent\begin{tabular}{@{}lcl@{}}
$f$         &=& E-sail thrust per unit tether length \\
$f_0$       &=& $f$ at $r=r_0$ \\
$\mathbf{F}$ &=& E-sail force acting on tether \\
$\mathbf{F}_s$ &=& Component of $\mathbf{F}$ along spin plane \\
$\mathbf{F}_n$ &=& Component of $\mathbf{F}$ perpendicular to spin plane \\
\textit{g}  &=& Factor by which maintether voltage is changed when auxtether segment is electrified \\
\textit{k}  &=& Factor by which centrifugal tension exceeds E-sail force \\
$l_\mathrm{aux}$ &=& Length of auxtether segment \\
\textit{m}  &=& Lumped mass of one tether subsystem (maintether, RemoteUnit and auxtether) \\
\textit{N}  &=& Number of tethers in the E-sail \\
\textit{r}  &=& Solar distance \\
$r_0$       &=& Base solar distance, 1 au \\
\textit{R}  &=& Tether length \\
\textit{t}  &=& Time \\
\textit{x}  &=& Cartesian coordinate along thrust vector component perpendicular to solar wind \\
\textit{y}  &=& Cartesian coordinate perpendicular to $x$ and $z$ so that $XYZ$ forms a right-handed system \\
\textit{z}  &=& Cartesian coordinate along solar wind direction \\
\textit{X,Y,Z} &=& Coordinate axes corresponding to coordinates $x,y,z$ \\
$\alpha$    &=& Angle between antisolarwind direction and E-sail spin axis \\
$\delta\varphi$ &=& Length of phase angle interval in which auxtether segment is electrified \\
$\tau_c$    &=& Coriolis torque acting on one tether subsystem \\
$\Delta\tau$&=& Instantaneous spin-changing torque generated by electrified auxtether segment\\
$\tau_\mathrm{ave}$ &=& Time average of $\Delta\tau$\\
{\boldmath$\omega$}    &=& E-sail spin angular frequency vector\\
$\Omega$    &=& Orbital angular frequency \\
$\Omega_0$  &=& Orbital angular frequency at $r=r_0$ \\
\end{tabular} \\

\section{Introduction}
\begin{figure}
\includegraphics[width=7.5cm]{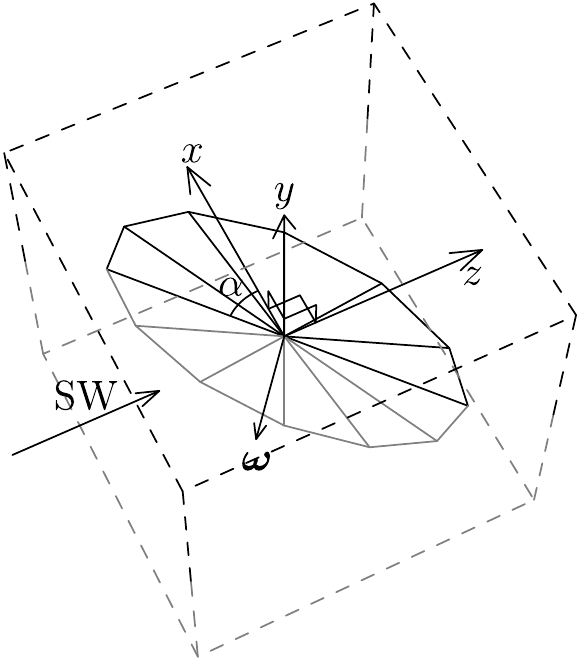}
\caption{Spinning E-sail inclined at angle $\alpha$ with respect to
  solar wind (SW) flow ($\alpha$ lies in the $XZ$ plane). Domain $y<0$ is drawn in grayscale to help visualization.}
\label{fig:3D}
\end{figure}

The electric solar wind sail (E-sail) uses charged tethers to tap
momentum from the solar wind (SW) \cite{paper1}. The baseline configuration
uses a set of radial main tethers which are kept under centrifugal
tension by spinning the system \cite{paper9}. For dynamical stability,
the tips of the main tethers are connected together by auxiliary
tethers \cite{paper9}. For each tether, the generated E-sail thrust
points along the tether-perpendicular component of the SW flow
direction. Thrust vectoring is thereby possible by inclining the sail
(the tether spin plane) with respect to the SW flow
(Fig.~\ref{fig:3D}). For each tether, the magnitude of the thrust can
be controlled by modulating the voltage of the tether, and sail
attitude maneuvers are possible by E-sail force only.

If the E-sail spacecraft orbits the Sun while having its sail
inclined, the orbital Coriolis acceleration interacts in a nontrivial
way with the spin of the tether rig \cite{paper14}. Specifically, the
spin rate accelerates (decelerates) if the sail is inclined so as to
spiral the orbit outwards (inwards) in the solar system. The angular
acceleration of the rotating sail is approximately given by
\begin{equation}
\frac{d\omega(t)}{dt} = \left(\Omega \tan\alpha\right) \omega(t)
\label{eq:angacc}
\end{equation}
where the sail inclination angle $\alpha$ is taken to be positive if
the sail accelerates away from the sun (i.e.~if the E-sail thrust
component along the orbital velocity vector is
positive). Equation \ref{eq:angacc} leads to exponential change of $\omega$
unless compensated by spin rate control.

The E-sail also needs spin rate control to generate angular momentum
during deployment of the tethers. Cold gas thrusters and ionic liquid
field effect electric propulsion (FEEP) thrusters have been prototyped
for spin rate control yielding RemoteUnit mass of 0.56 kg \cite{D412},
and solar photonic blades have been considered at lower technical
readiness level (TRL) \cite{paper16}. Here we propose a new intrinsic
method of controlling the spin rate by the E-sail force itself.

\section{Spin rate modulation by electrified auxtethers}

Consider an E-sail as in Fig.~\ref{fig:3D} where now the auxiliary
tethers (auxtethers) can at least partly be put under voltage. A
segment of auxtether generates E-sail thrust which is perpendicular to
it if one neglects end effects. We now want to show that if the auxtether
voltage can be controlled independently from the adjacent maintether voltage,
pure E-sail spin rate control becomes possible.

\begin{figure}
\includegraphics[width=14cm]{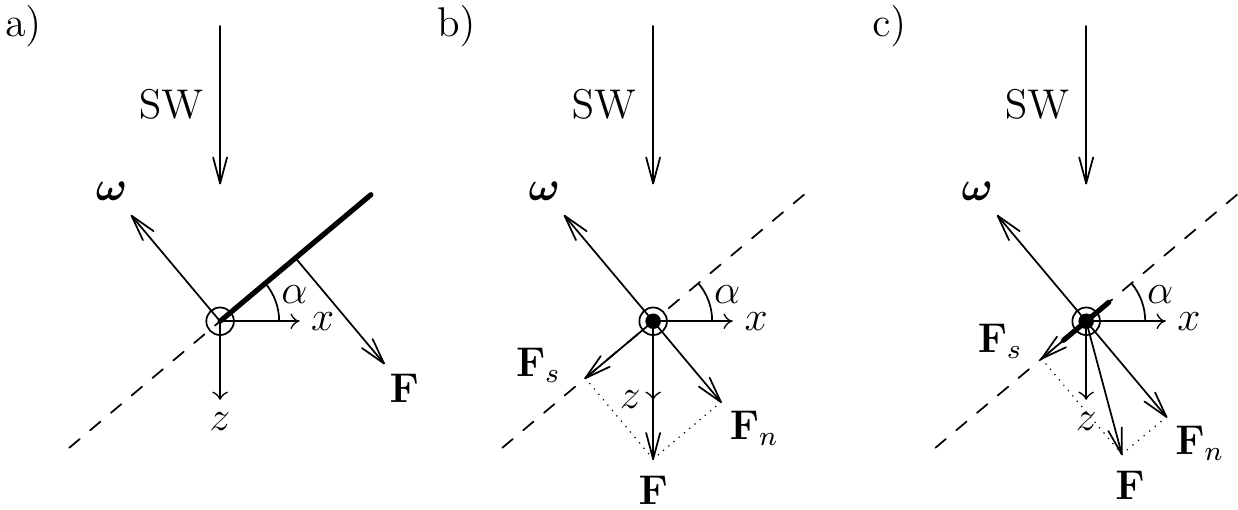}
\caption{(a) Considering a single tether (thick line) lying momentarily in the $XZ$ plane and viewed
  along $Y$. E-sail force $\mathbf{F}$ is along the tether-perpendicular component
  of the SW. (b) When tether is parallel to $Y$, it is already
  perpendicular to SW and $\mathbf{F}$ is aligned with
  SW. $\mathbf{F}$ is decomposed to spinplane aligned component $\mathbf{F}_s$ and
  spinplane perpendicular component $\mathbf{F}_0$. (c)
  Tether parallel to $Y$ and with adjacent electrified auxtether
  segment. Spinplane E-sail force component $\mathbf{F}_s$ is smaller
  than in (b).}
\label{fig:Principle}
\end{figure}

Figure \ref{fig:Principle}a shows an E-sail inclined at angle $\alpha$
to the SW flow and one of its charged maintethers which at the moment
happens to lie in the $XZ$ plane i.e.~in the plane of the
figure. At this moment the maintether in question generates thrust
vector $\mathbf{F}$ which is perpendicular to the tether spinplane
(dashed line), i.e.~inclined by angle $\alpha$ with respect to the SW.

Figure \ref{fig:Principle}b shows the same maintether a quarter cycle
later when it is aligned along $Y$ axis. Now the maintether is already
perpendicular to the SW so that the thrust $\mathbf{F}$ is
simply aligned with the SW. $\mathbf{F}$ can be decomposed
into spinplane component $\mathbf{F}_s$ and
spinplane normal component $\mathbf{F}_n$. The spinplane
component of the E-sail thrust tends to decrease the tether's $\omega$
when it climbs upstream. The reverse happens half a cycle later when
the tether moves downstream so that overall no change in spin rate is
generated.

In Fig.~\ref{fig:Principle}c there is a short charged auxtether
segment attached to the tip of the maintether at the same moment as in
\ref{fig:Principle}b. The E-sail thrust vector $\mathbf{F}$ is now a
vector sum of the maintether thrust and the auxtether thrust. The
maintether thrust is along the SW flow as in \ref{fig:Principle}b, but
the auxtether thrust is perpendicular to the auxtether,
i.e.~perpendicular to the spin plane. As a result, $\mathbf{F}$ is not
fully aligned with the SW and its spinplane component $F_s$ is
smaller. To prevent the sail from turning, it is necessary that $F_n$
has the same value when the tether is moving upstream and
downstream. However, the spinplane component $F_s$ can be different
because the same $F_n$ can be obtained by different combinations of
the maintether and auxtether voltages.  Effectively, we have two
control parameters (maintether voltage and auxtether voltage) which
allows us to control the two thrust parameters $F_s$ and $F_n$
freely. By having the same $F_n$ but different $F_s$ in the upstream
and downstream portions of the maintether's cycle, one can modify the
sail's spin rate while keeping its orientation fixed.

\section{Effect strength}

Consider an E-sail consisting of $N$ maintethers of length $R$,
approximating each by a point mass $m$ located at the tip so that $m$
contains the masses of the maintether, the Remote Unit and the
associated auxtether segment. Assume that the E-sail thrust per unit
length $f$ at solar distance $r_0$ (select $r_0 = 1$ au for
definiteness) is $f_0$. At other distances $r$, $f =
f_0(r_0/r)$ \cite{paper9}. Assume also that at each solar distance $r$
the spin rate $\omega$ is chosen such that the centrifugal tension of
the maintether is $k$ times the E-sail force acting on it where one
typically requires $k\gtrsim 5$ for stability. This implies $m R
\omega^2 = k f R$ so that
\begin{equation}
\omega = \sqrt{\frac{k f}{m}} = \sqrt{\frac{k f_0 r_0}{m r}}.
\label{eq:omega}
\end{equation}
Equation (\ref{eq:angacc}) gives the secular angular acceleration due to
the orbital Coriolis effect. The corresponding Coriolis torque is
\begin{equation}
\tau_c = m R^2 \frac{d\omega}{dt} = m R^2 \Omega \omega \tan\alpha.
\label{eq:tauc}
\end{equation}
Thus, Eq.~(\ref{eq:tauc}) gives the torque that must be compensated by
the electrified auxtether E-sail effect.

Consider the tether when it's pointed along $Y$
(Fig.~\ref{fig:Principle}b,c) where the length of its associated
auxtether segment is $l_\mathrm{aux} = 2\pi R/N$. If the maintether
has voltage and the auxtether has not, the E-sail thrust acting on the
maintether is $F = f R$ whose components are $F_s = F \sin\alpha$ and
$F_n = F \cos\alpha$. If both tethers have voltage, then $F_s = g f
R \sin\alpha$ and $F_n = g f R \cos\alpha + f l_\mathrm{aux}$ where
$g$ is a dimensionless factor by which the maintether voltage is
modulated ($0\le g\le 1$). Requiring that $F_n$ causes the same torque in both cases implies
\begin{equation}
1-g = \frac{2 l_\mathrm{aux}}{R\cos\alpha}
\label{eq:g}
\end{equation}
and the difference in spin axis aligned torques becomes
\begin{equation}
\Delta\tau = f l_\mathrm{aux} R \tan\alpha = \frac{2\pi}{N} f R^2 \tan\alpha.
\label{eq:deltatau}
\end{equation}


Let us assume that the auxtether voltage is used only when the tether
is close to vertical, in phase angle interval of length $\delta\varphi$, so that the
average E-sail torque due to auxtether electrification is
$\tau_\mathrm{ave} = (\delta\varphi/(2\pi))\Delta\tau$. Requiring
$\tau_\mathrm{ave} = \tau_c$ we can solve for $\delta\varphi$. Using
above definitions, the result can be expressed as
\begin{equation}
\delta\varphi = Nk\frac{\Omega}{\omega}
= N \sqrt{k} \left(\frac{r_0}{r}\right) \Omega_0 \sqrt{\frac{m}{f_0}}
\end{equation}
where $\Omega_0$ is the orbital angular frequency at $r=r_0$:
$\Omega=\Omega_0(r_0/r)^{3/2}$. For example for the baseline 1 N
E-sail orbiting at 1 au $N=100$, $k=5$, $r=r_0$,
$\Omega_0=2\pi/\mathrm{year} = 2.0\cdot 10^{-7}$ s$^{-1}$, $m=1$ kg,
and $f_0=500$ nN/m so that $\delta\varphi = 3.6^\circ$. This is a
conveniently small fraction of the full circle, hence there is
typically an abundant performance margin that can be used e.g.~for
running the auxtethers at reduced voltage or for having only part of
the auxtether being conductive.


\section{Conclusion}
In many missions the orbital Coriolis effect causes a secular
variation of the E-sail spin rate which must be compensated by
technical means. We showed conceptually that this can be effected by making the
auxtethers partly or completely from conducting material and
having a way to control their voltages. The resulting pure E-sail concept
does not need any auxiliary propulsion systems such as cold gas
thrusters, FEEP thrusters or photonic blades for its long-term spin rate management.

In the future, more refined engineering studies should be carried out
concerning conducting auxtether designs and the voltage regulation
mechanisms of the auxtethers. One should also study the details of how
the proposed scheme can be used for angular momentum generation during
tether deployment.

\section*{Acknowledgments}
The work was funded by the Academy of Finland grant 250591.


\section*{References}

\begin{thebibliography}{}

\bibitem{paper1}
Janhunen, P., ``Electric sail for spacecraft propulsion,''
\textit{J.~Prop.~Power}, Vol.~20, No.~4, 2004, pp.~763, 764. doi: 10.2514/1.8580

\bibitem{paper9}
Janhunen, P., et al., ``Electric solar wind sail: Toward test
missions,'' \textit{Rev.~Sci.~Instrum.}, Vol.~81, No.~11, 2010,
pp.~111301. doi: 10.1063/1.3514548

\bibitem{paper14}
Toivanen, P.K., and Janhunen, P., ``Spin plane control and thrust
vectoring of electric solar wind sail by tether potential
modulation,'' \textit{J.~Prop.~Power}, Vol.~29, No.~1, 2013, pp.~178,
185. doi: 10.2514/1.B34330

\bibitem{D412}
Wagner, S., Sundqvist, J., and Thornell, G., ``Design description of the
Remote Unit,'' \textit{ESAIL FP7 project deliverable D41.2},
2012. Available at {\tt
  http://www.electric-sailing.fi/fp7/docs/D412.pdf} 

\bibitem{paper16}
Janhunen, P., ``Photonic spin control for solar wind electric sail,''
\textit{Acta Astronaut.}, Vol.~83, 2013, pp.~85, 90. doi:
10.1016/j.actaastro.2012.10.017


\end{thebibliography}
\end{document}